\documentclass{emulateapj}

\usepackage{amssymb}

\def\ts     {\thinspace}
\def\kms    {\ts km\ts s$^{-1}$}
\def\etal   {{\rm et\ts al.}}
\def\msol   {$M_{\odot}$}
\def\lsol   {$L_{\odot}$}
\def\lprime {K\,\kms\,pc$^2$}

\def\aco    {{\rm CO}($J$=1$\to$0)}

\def\cco    {{\rm CO}($J$=3$\to$2)}

\slugcomment{draft version \today, accepted for publication in the Astrophysical Journal Letters}

\shorttitle{\cco\ imaging of a Submillimeter-Selected $z$=2.8 Quasar}
\shortauthors{Riechers}

\begin{document}

\title{
SMM J04135+10277:\\ A Candidate Early--Stage ``Wet--Dry'' Merger of Two Massive Galaxies at $z$=2.8}

\author{Dominik A.\ Riechers\altaffilmark{1,2}}

\altaffiltext{1}{Astronomy Department, California Institute of
  Technology, MC 249-17, 1200 East California Boulevard, Pasadena, CA
  91125, USA}
\altaffiltext{2}{Astronomy Department, Cornell University, 220 Space Sciences Building, Ithaca, NY 14853, USA; dr@astro.cornell.edu}


\begin{abstract}

  We report interferometric imaging of \cco\ emission toward the
  $z$=2.846 submillimeter-selected galaxy SMM\,J04135+10277, using the
  Combined Array for Research in Millimeter-wave Astronomy (CARMA).
  SMM\,J04135+10277 was previously thought to be a gas-rich,
  submillimeter-selected quasar, with the highest molecular gas mass
  among high-$z$ quasars reported in the literature.  Our maps at
  $\sim$6$\times$ improved linear resolution relative to earlier
  observations spatially resolve the emission on $\sim$1.7$''$ scales,
  corresponding to a (lensing-corrected) source radius of
  $\sim$5.2\,kpc. They also reveal that the molecular gas reservoir,
  and thus, likely the submillimeter emission, is not associated with
  the host galaxy of the quasar, but with an optically faint gas-rich
  galaxy at 5.2$''$, or 41.5\,kpc projected distance from the active
  galactic nucleus (AGN).  The obscured gas-rich galaxy has a
  dynamical mass of $M_{\rm dyn}\,{\rm
    sin}^2i$=5.6$\times$10$^{11}$\,\msol, corresponding to a gas mass
  fraction of $\simeq$21\%.  Assuming a typical $M_{\rm BH}$/$M_*$
  ratio for $z$$\gtrsim$2 quasars, the two galaxies in this system
  have an approximate mass ratio of $\sim$1.9.  Our findings suggest
  that this quasar--starburst galaxy pair could represent an early
  stage of a rare major, gas-rich/gas-poor (``wet--dry'') merger of
  two massive galaxies at $z$=2.8, rather than a single, gas-rich AGN
  host galaxy.  Such systems could play an important role in the early
  buildup of present-day massive galaxies through a
  submillimeter-luminous starburst phase, and may remain hidden in
  larger numbers among rest-frame far-infrared-selected quasar samples
  at low and high redshift.

\end{abstract}


\section{Introduction}

Studies of gas- and dust-rich, starbursting AGN host galaxies out to
cosmological distances are important to better understand the
connection between supermassive black hole and stellar bulge growth in
galaxies that gives rise to the present-day $M_{\rm BH}$--$M_{\rm
bulge}$ relation (Magorrian et al.\ \citeyear{mag98}; H\"aring \& Rix
\citeyear{hr04}). A particularly important cosmic epoch for these
studies is the redshift range 2$\lesssim$$z$$\lesssim$3 where most of
the growth of stellar and black hole mass in galaxies occurs, i.e.,
where the volume densities of both cosmic star formation and AGN
activity peak (e.g., Magnelli et al.\ \citeyear{mag09}; Richards et
al.\ \citeyear{ric06}).

It has recently been found that the dynamical masses of some of the
most distant quasars at $z$$>$4 appear to be too small to host stellar
components as expected from the local $M_{\rm BH}$--$M_{\rm bulge}$
relation, and that the available gas masses are too small to produce a
sufficient amount of stars to approach this relation (e.g., Walter et
al.\ \citeyear{wal04}; Riechers et al.\ \citeyear{rie08a},
\citeyear{rie08b}). Thus, these galaxies appear to require a source of
external gas supply (or stars) to assemble sufficient stellar mass by
$z$=0 to approach the local $M_{\rm BH}$--$M_{\rm bulge}$ relation.
This gas supply could either be due to accretion of gas through cold
streams (e.g., Dekel et al.\ \citeyear{dek09}), or due to gas-rich,
gas-rich (``wet-wet'') or gas-rich, gas-poor (``wet-dry'') mergers
with massive and/or gas-rich galaxies (e.g., Springel et al.\
\citeyear{spr05}).

Examples of the latter may be found among high-redshift,
submillimeter-selected quasars. A strong submillimeter detection is
suggestive of a large amount of warm dust heated by young stars formed
at high rates (e.g., Isaak et al.\ \citeyear{isa02}). Follow-up
observations of the molecular interstellar medium (ISM) in these
galaxies, typically through the detection of CO lines, are important
to measure the mass of the ISM that constitutes the reservoir for star
formation, and to confirm that the starburst and gas are at the same
redshift as the AGN (e.g., Coppin et al.\ \citeyear{cop08}).

A particularly interesting submillimeter-selected quasar was found in
the field of the $z$=0.088 galaxy cluster Abell\,478,
SMM\,J04135+10277 at $z$=2.837$\pm$0.003 (Knudsen et al.\
\citeyear{knu03}). The source was identified in 450 and 850\,$\mu$m
observations with the JCMT/SCUBA instrument, revealing high
submillimeter fluxes of 25$\pm$2.8 and 55$\pm$17\,mJy, respectively,
which suggest a total infrared luminosity of
(2.9$\pm$0.5)$\times$10$^{13}$\,\lsol. Subsequent interferometric
\cco\ observations at 15$''$$\times$11$''$ resolution and single-dish
\aco\ observations revealed a massive molecular gas reservoir at
$z$=2.846$\pm$0.002, consistent with both the redshift and position of
the quasar within the relative uncertainties (Hainline et al.\
\citeyear{hai04}; Riechers et al.\ \citeyear{rie11a}). None of these
past studies offered sufficient spatial resolution to spatially
resolve and/or precisely locate the CO or submillimeter continuum
emission.

We here report higher spatial resolution \cco\ observations with CARMA
to determine the size and dynamical mass of the molecular gas
reservoir. We use a concordance, flat $\Lambda$CDM cosmology
throughout, with $H_0$=71\,\kms\,Mpc$^{-1}$, $\Omega_{\rm M}$=0.27,
and $\Omega_{\Lambda}$=0.73 (Spergel \etal\
\citeyear{spe03}, \citeyear{spe07}).

\section{Observations}

\begin{figure}
\epsscale{1.15}
\plotone{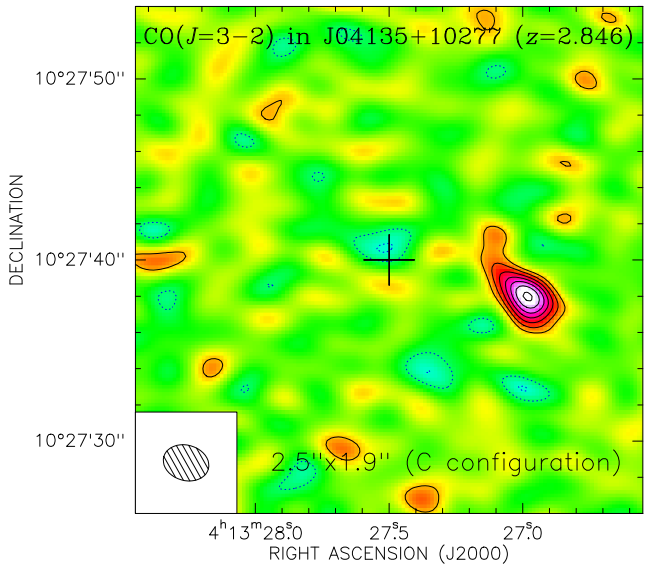}

\caption{CARMA \cco\ map of SMM\,J04135+10277 over the central 765\,\kms. 
Contours are shown in steps of 1$\sigma$=0.555\,mJy\,beam$^{-1}$,
starting at $\pm$2$\sigma$. The cross indicates the pointing
center. The synthesized beam size is shown in the lower left corner.
\label{f1}}
%
\end{figure}

We observed the \cco\ transition line ($\nu_{\rm rest} =
345.7959899\,$GHz, redshifted to 89.911\,GHz, or 3.33\,mm) towards
SMM\,J04135+10277, using CARMA.  A total bandwidth of 3.7\,GHz
($\sim$12,400\,\kms; at 5.208\,MHz resolution) was used to cover the
\cco\ line and the underlying 3.33\,mm (rest-frame 870\,$\mu$m)
continuum emission.  Observations were carried out under good 3\,mm
weather conditions for 3\,tracks in C configuration (18--367\,m
baselines, which corresponds to probing spatial scales of
1.5$''$--31$''$, or 11--250\,kpc) on 2012 February 25, 29, and March
12.  This resulted in 6.3\,hr of 15 antenna-equivalent on-source time
after discarding unusable visibility data. The nearby source 3C120 was
observed every 15 minutes for pointing, amplitude and phase
calibration. Fluxes were bootstrapped relative to Mars. The bright
nearby calibrators 3C84 and J0423--013 were observed for bandpass
calibration, yielding $\sim$15\% calibration accuracy.

The MIRIAD package was used for data reduction and analysis.  All data
were mapped using the CLEAN algorithm with `natural' weighting,
resulting in a synthesized beam size of 2.5\,$''$$\times$1.9\,$''$.
The final rms is 0.55\,mJy beam$^{-1}$ over 229.2\,MHz (corresponding
to 765\,\kms), and 1.8\,mJy beam$^{-1}$ over 20.8\,MHz (69\,\kms).

\section{Results}

\begin{figure}
\epsscale{1.15}
\plotone{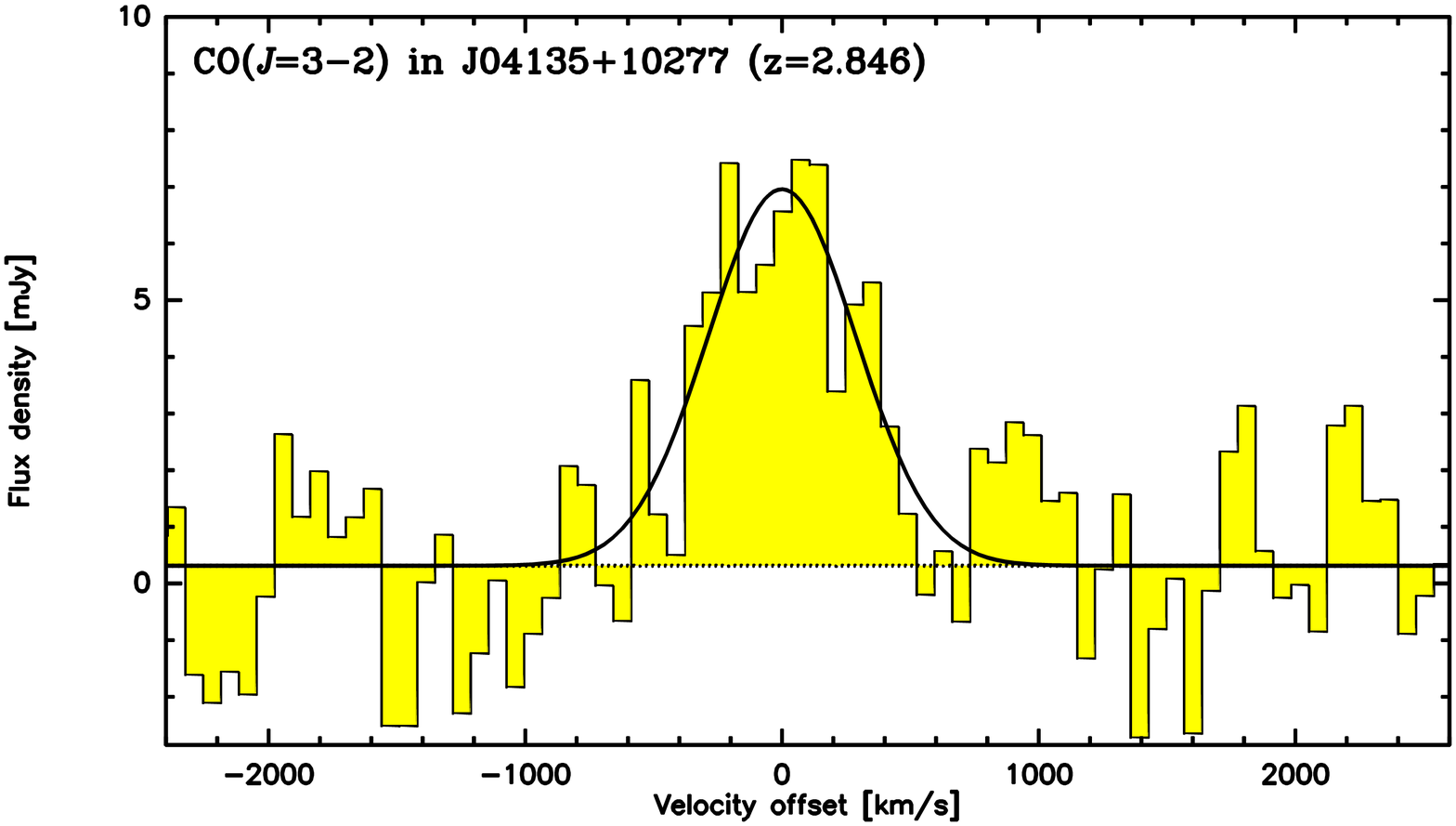}

\caption{Spectrum of \cco\ emission toward SMM\,J04135+10277. The spectrum (histogram) is shown at 20.8\,MHz (69\,\kms ) resolution. The solid curve indicates a Gaussian fit to the spectrum. \label{f2}}
%
\end{figure}

We have detected and spatially resolved strong \cco\ emission toward
SMM\,J04135+10277 (Fig.~\ref{f1}). CO emission is detected in each of
the tracks individually at the same position. By fitting an
elliptical, two-dimensional Gaussian to the $u-v$ data, we find a
source size of 1.66$''$$\pm$0.40$''$ along its major axis,
corresponding to 13.2$\pm$3.2\,kpc at $z$=2.846. The source remains
unresolved down to $\lesssim$1.2$''$ ($\lesssim$9.5\,kpc) along its
minor axis. From fitting a four-parameter Gaussian to the spectrum of
the \cco\ line emission (Fig.~\ref{f2}), we measure a line peak flux
of 6.6$\pm$0.9\,mJy at a FWHM velocity width of 679$\pm$120\,\kms.
Within the relative uncertainties, the line width is consistent with
that measured in the \aco\ line (505$\pm$75\,\kms; Riechers et al.\
\citeyear{rie11a}), and marginally consistent with a previous
measurement of the \cco\ line at lower significance
(340$\pm$120\,\kms; Hainline et al.\ \citeyear{hai04}). Our
measurements correspond to an integrated \cco\ line flux of
4.78$\pm$0.67\,Jy\,\kms, and a \cco/\aco\ brightness temperature ratio
of $r_{31}$=0.82$\pm$0.15. The Gaussian peaks at a redshift of $z_{\rm
  CO}$=2.8458$\pm$0.0006, consistent with previous estimates within
the errors. We marginally detect the underlying 3.33\,mm continuum
emission at a level of 0.31$\pm$0.17\,mJy.

\begin{figure}
\vspace{1.5mm}
\epsscale{1.15}
\plotone{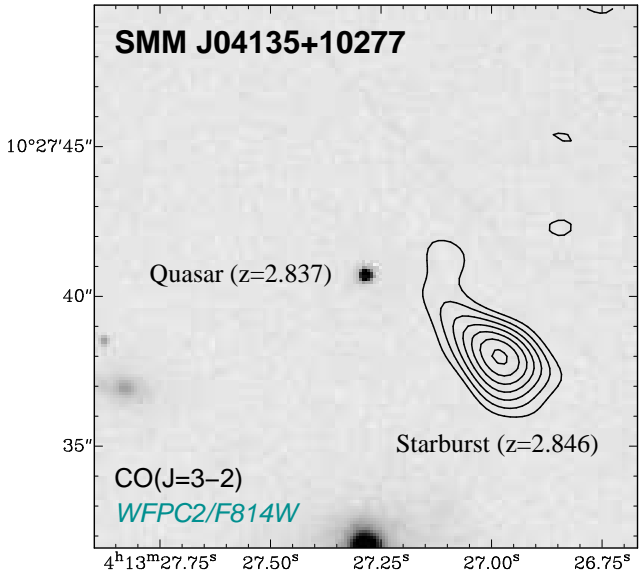}

\caption{Overlay of \cco\ emission (contours) toward SMM\,J04135+10277 on a {\em Hubble Space Telescope} WFPC2 F814W image (rest-frame $\sim$210\,nm). The CO emission is spatially offset by 5.2$''$ from the position of the optical quasar. No optical emission is detected at the position of the CO emission.\label{f3}}
%
\end{figure}

\begin{figure*}
\epsscale{1.15}
\plotone{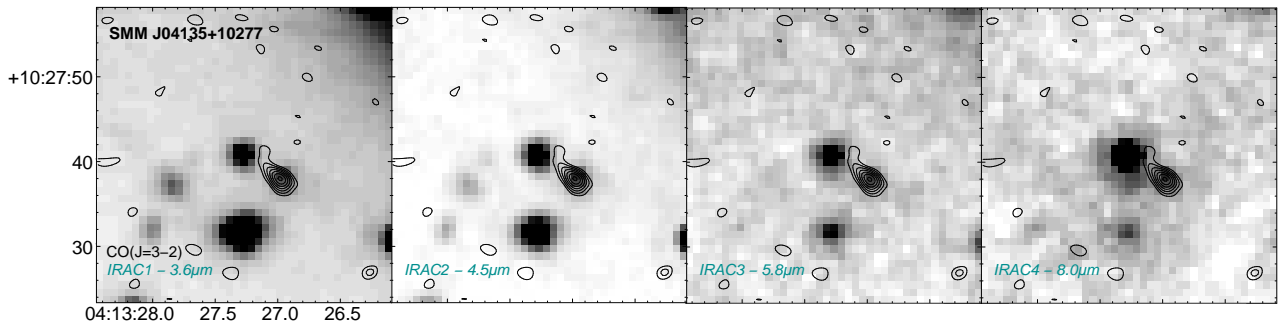}

\caption{Overlay of \cco\ emission (contours) toward SMM\,J04135+10277 on {\em Spitzer Space Telescope} IRAC band 1--4 images (rest-frame $\sim$0.94--2.08\,$\mu$m). Faint emission is detected at the position of the CO emission in all IRAC bands, with a flatter spectral slope than that of the optical quasar.\label{f4}}
%
\end{figure*}

The \cco\ emission peaks at a position of $\alpha$=$04^{\rm h}13^{\rm
  m}26^{\rm s}.989$ $\pm$ $0''.11$, $\delta$=$+10^\circ27'37''.89$
$\pm$ $0''.12$. The $i$-band position of the quasar as determined from
a {\em Hubble Space Telescope} WFPC2 F814W image obtained from the
Hubble Legacy Archive\footnote{\tt http://hla.stsci.edu} is
$\alpha$=$04^{\rm h}13^{\rm m}27^{\rm s}.28$,
$\delta$=$+10^\circ27'40''.77$. Thus, the molecular gas reservoir is
spatially offset by 5.2$''$, or 41.5\,kpc, from the AGN position
(Fig.~\ref{f3}). There is no evidence for any rest-frame $\sim$210\,nm
emission at the position of the CO emission. {\em Spitzer Space
  Telescope} IRAC 3.6--8.0\,$\mu$m images obtained from the Spitzer
Heritage Archive\footnote{\tt http://sha.ipac.caltech.edu} reveal a
faint counterpart to the CO-emitting galaxy at rest-frame
near-infrared wavelengths (0.9--2.1\,$\mu$m; Fig.~\ref{f4}). Its
rest-frame near-infrared spectrum appears to be flatter than that of
the quasar, consistent with a dust-obscured star-forming galaxy and
the lack of a strong active galactic nucleus component.  There is no
evidence for any CO or continuum emission at the position of the
quasar.  Assuming a line FWHM of 400\,\kms, we derive a 3$\sigma$
upper limit of 0.9\,Jy\,\kms\ for the \cco\ line flux of the quasar
host galaxy.  This corresponds to $<$20\% of the \cco\ line flux of
the submillimeter source. We also derive a 3$\sigma$ upper limit of
0.4\,mJy for the 3.33\,mm continuum emission at the position of the
quasar.

\section{Analysis}

To account for a small amount of gravitational magnification by the
foreground galaxy cluster, we will adopt a lensing magnification
factor of $\mu_{\rm L}^{\rm QSO}$=1.3 for the active galactic nucleus
(as determined by Knudsen et al.\ \citeyear{knu03}), and of $\mu_{\rm
  L}^{\rm CO}$=1.6$\pm$0.5 for the molecular gas and far-infrared
continuum emission (as determined using the CO $J$=1$\to$0 line
luminosity and FWHM from Riechers et al.\ \citeyear{rie11a} and a
$\mu_{\rm L}$--$L'_{\rm CO(1-0)}$--d$v_{\rm FWHM}$ scaling relation
for submillimeter-selected galaxies; Harris et al.\ \citeyear{har12})
in the following. We further adopt the gas mass determined by Riechers
et al.\ (\citeyear{rie11a}) based on the \aco\ line luminosity, but
scaled to our updated lensing magnification factor.  This suggests
$M({\rm H_2}) = 1.2 \times 10^{11}$\,(1.6/$\mu_{\rm
  L}$)$^{-1}$\,M$_{\odot}$ (assuming a $L'_{\rm CO(1-0)}$--$M({\rm
  H_2})$ conversion factor of $\alpha_{\rm CO}$=0.8\,\msol\,(\lprime
)$^{-1}$ for ultra-luminous infrared galaxies (ULIRGs); Downes \&
Solomon \citeyear{ds98}, but also see recent discussion by
Papadopoulos et al.\ \citeyear{pap12}). The size estimate (5.2\,kpc
radius; corrected by a factor of $\mu_{\rm L}^{1/2}$; e.g., Riechers
et al.\ \citeyear{rie09a}) and width of the CO line suggest a
dynamical mass of $M_{\rm dyn}\,{\rm
  sin}^2i$=5.6$\times$10$^{11}$\,\msol.\footnote{Due to the flattening
  of the baryonic mass distribution in a disk, possible biases due to
  clumpiness of the gas, and non-circular motions of the gas, virial
  estimates for clumpy, disk-like galaxies may underpredict the
  dynamical mass by typically $\sim$30\% (e.g., Daddi et al.\
  \citeyear{dad10a}). Estimates for more complex dynamical systems
  likely exhibit at least comparable uncertainties, and thus, require
  model-based correction factors. For submillimeter-bright dusty
  starburst galaxies, an isotropic virial estimator is commonly
  adopted, which gives by a factor of $\sim$1.5 larger values than
  standard estimates for rotating disk galaxies at an average
  inclination (Engel et al.\ \citeyear{eng10}). Adopting the isotropic
  virial estimator instead of the inclined disk model used here would
  result in a $\sim$20\% larger dynamical mass relative to an edge-on
  configuration. Both estimators result in the same value when
  assuming an inclination of $i$$\simeq$66$^\circ$. The uncertainties
  on the dynamical mass estimate, and thus, $f_{\rm gas}$, amount to
  at least 20\%--30\%.}  This suggests a gas mass fraction of $f_{\rm
  gas}$=$M({\rm H_2})$/($M_{\rm dyn}\,{\rm sin}^2i$)$\simeq$21\%.

To determine the black hole mass of the quasar, we adopt the
$\lambda$$L_{5100\AA }$ luminosity determined by Knudsen et al.\
(\citeyear{knu03}) and the $M_{\rm BH}$--$L_{5100\AA }$ relationship
determined by Peterson et al.\ (\citeyear{pet04}). This suggests a
black hole mass of $M_{\rm BH}$=1.7$\times$10$^9$\,\msol. If the
source were to follow the $M_{\rm BH}$--$M_{\rm bulge}$ relation for
nearby galaxies (H\"aring \& Rix \citeyear{hr04}), this would suggest
$M_{\rm bulge}$$\simeq$8.2$\times$10$^{11}$\,\msol.  However, Peng et
al.\ (\citeyear{pen06}) suggest that this ratio is likely typically
$\sim$4$\times$ lower at $z$$\sim$2. In the following, we thus assume
a stellar mass of $M_\star$$\simeq$2$\times$10$^{11}$\,\msol\ for the
quasar host galaxy. Approximating the total mass of the system as
$M_{\rm tot}$=$M_{\rm BH}$+$M_\star$+$M({\rm H_2})$+$M_{\rm
  dust}$+$M_{\rm DM}$$\simeq$3$\times$10$^{11}$\,\msol\ (adopting a
gas-to-dust ratio of 100, and assuming a contribution from dark matter
(DM) of 25\%), we find a gas fraction of $f_{\rm gas}^{\rm t}$=$M({\rm
  H_2})$/$M_{\rm tot}$$<$8\% (assuming the same gas excitation as for
the CO-detected galaxy).

\section{Discussion and Conclusions}

We have imaged \cco\ emission towards the high-redshift galaxy
SMM\,J04135+10277, using CARMA. Our observations suggest that the
molecular gas reservoir previously detected towards SMM\,J04135+10277
is not associated with the host galaxy of the optically-detected
quasar at $z$=2.837, but with an optically faint, gas-rich galaxy at
$z$=2.846, separated by $\gtrsim$40\,kpc from the active galactic
nucleus. The spectral properties of the CO-emitting source are
consistent with those of an optically obscured star-forming galaxy.
Its \cco/\aco\ brightness temperature ratio of $r_{31}$=0.82$\pm$0.15
is within the range of values observed for $z$$>$2
submillimeter-selected starburst galaxies (e.g., Riechers et al.\
\citeyear{rie11c}, \citeyear{rie11d}; Ivison et al.\
\citeyear{ivi11}), but higher than the values typically observed in
high-$z$ disk galaxies (e.g., Dannerbauer et al.\ \citeyear{dan09};
Aravena et al.\ \citeyear{ara10}).  Assuming no difference in CO
excitation, this gas-rich companion carries at least 5$\times$ the gas
mass of the quasar host galaxy.\footnote{High redshift quasar host
  galaxies commonly show high $r_{31}$ of $>$0.9, which may suggest an
  even higher ratio in gas mass (e.g., Riechers et al.\
  \citeyear{rie06}, \citeyear{rie11a}).} The gas mass and gas mass
fraction of the companion are comparable to those of other $z$$>$2
submillimeter galaxies (e.g., Tacconi et al.\ \citeyear{tac06};
Riechers et al.\ \citeyear{rie11b}).  We estimate the total mass of
the SMM\,J04135+10277 system to be
$\gtrsim$8.6$\times$10$^{11}$\,\msol\ (with considerable uncertainty),
which may be dominated by the gas-rich companion galaxy (mass ratio of
$\sim$1.9).

Gas-rich companions have been detected in other high-$z$ quasars as
well, such as the $z$=4.4 and 4.7 systems BRI\,1335--0417 (which is
already actively merging with its companion on 5\,kpc scales; Riechers
et al.\ \citeyear{rie08a}) and BR\,1202--0725 (which has a companion
at 26\,kpc projection; e.g., Carilli et al.\ \citeyear{car02},
\citeyear{car13}). However, in all examples known so far at high $z$,
both the quasar host and companion are gas-rich.  Thus, the
quasar--starburst galaxy pair SMM\,J04135+10277 could be the first
high redshift example of an early-stage gas-rich, gas-poor merger, in
which the optically faint submillimeter galaxy provides the gas supply
to further build up the stellar component of the quasar host.

Models of hierarchical structure formation (e.g., Springel et al.\
\citeyear{spr05}) lend support to the idea that a close massive galaxy
pair like SMM\,J04135+10277 will likely result in a major merger in
which the gas-rich companion may replenish the gas supply in the
quasar host (which perhaps will yield configurations similar to those
observed in nearby infrared-luminous galaxies in intermediate or late
merger stages that contain both AGN and gas-rich galaxy components;
e.g., Evans et al.\ \citeyear{eva02}). The relatively large projected
separation of $\gtrsim$40\,kpc suggests that the two massive galaxies
are likely physically related and gravitationally interacting, but
cannot (yet) be considered part of a common gravitational potential.
This is consistent with what is expected for a merging system in an
early stage.  Given the early phase in the merging process implied by
this scenario, it is plausible but not unambiguous to assume that the
ongoing black hole accretion in the quasar and star formation in the
gas-rich companion could have been triggered by interaction.

Submillimeter-selected high redshift quasars are good candidates for
transition objects from hyper-luminous infrared galaxies to optically
bright quasars, linking the most intense starbursts in the universe
the the most actively accreting black holes (e.g., Coppin et al.\
\citeyear{cop08}; Simpson et al.\ \citeyear{sim12}). This scenario
would be consistent with a high redshift analogue of the ULIRG--quasar
transition scenario proposed by Sanders et al.\ (\citeyear{san88}).
However, despite being a good candidate for a transition object
initially, SMM\,J04135+10277 is not an example of such sources.
Instead, our observations suggest that it is a good candidate for an
early-stage gas-rich, gas-poor (``wet-dry'') merger\footnote{We note
  that ``dry'' here refers to the fact that the companion galaxy far
  dominates the gas content of this system. It does not imply that
  there is no gas whatsoever in the host of the quasar.} of two
massive galaxies at $z$=2.8, and thus, a possibly more extreme
high-redshift analogue to the $z$=0.3 quasars HE\,0450--2958 and
J1821+643 (which were identified to not be transition objects through
similar observational strategies; e.g., Papadopoulos et al.\
\citeyear{pap08}; Aravena et al.\ \citeyear{ara11}).

Gas-rich, gas-poor (``wet-dry'') mergers at high redshift are
predicted by cosmological simulations of hierarchical structure
formation (e.g., Springel et al.\ \citeyear{spr05}), but successful
observations of such systems are still scarce. Based on the discovery
of SMM\,J04135+10277 alone, it remains unclear what the incidence of
such systems within submillimeter-selected quasar samples is. It
however clearly motivates observations of larger samples with
high-quality optical/infrared data in CO and submillimeter continuum
emission at high spatial resolution. Such studies will become feasible
in the near future, with the completion of both the Karl G.\ Jansky
Very Large Array (VLA) and the Atacama Large (sub-) Millimeter Array
(ALMA).

\acknowledgments

We thank the anonymous referee for a helpful report. DR acknowledges
support from the National Aeronautics and Space Administration (NASA)
through a Spitzer Space Telescope grant. Support for CARMA
construction was derived from the G.\ and B.\ Moore Foundation, the
K.~T.\ and E.~L.\ Norris Foundation, the Associates of the California
Institute of Technology, the states of California, Illinois, and
Maryland, and the NSF. Ongoing CARMA development and operations are
supported by the NSF under a cooperative agreement, and by the CARMA
partner universities. Based in part on observations made with the
NASA/ESA Hubble Space Telescope, and obtained from the Hubble Legacy
Archive, which is a collaboration between the Space Telescope Science
Institute (STScI/NASA), the Space Telescope European Coordinating
Facility (ST-ECF/ESA), and the Canadian Astronomy Data Centre
(CADC/NRC/CSA). Based in part on observations made with the Spitzer
Space Telescope, and obtained from the Spitzer Heritage Archive
through the NASA/IPAC Infrared Science Archive, which is operated by
the Jet Propulsion Laboratory, California Institute of Technology,
under contract with NASA.

\end{document}